\documentclass[%
 preprint,
 amsmath,amssymb,
 aps,
 superscriptaddress,
]{revtex4-2}

\usepackage{graphicx}
\usepackage{dcolumn}
\usepackage{bm}
\usepackage[utf8]{inputenc}
\usepackage{amssymb}
\usepackage{amsmath}
\usepackage{bm}
\usepackage{color}
\usepackage{url}
\usepackage{placeins}
\usepackage{array}
\newcolumntype{P}[1]{>{\centering\arraybackslash}p{#1}}
\newcolumntype{M}[1]{>{\centering\arraybackslash}m{#1}}

\begin{document}


\title{The effects of local homogeneity assumptions in metapopulation models of infectious disease}

\author{Cameron Zachreson}
\affiliation{School of Computing and Information Systems, The University of Melbourne, Australia}
\affiliation{Centre for Complex Systems, Faculty of Engineering,
The University of Sydney, Sydney, New South Wales, 2006, Australia}
\author{Sheryl Chang}
\author{Nathan Harding}
\affiliation{Centre for Complex Systems, Faculty of Engineering,
The University of Sydney, Sydney, New South Wales, 2006, Australia}
\author{Mikhail Prokopenko}
\affiliation{Centre for Complex Systems, Faculty of Engineering,
The University of Sydney, Sydney, New South Wales, 2006, Australia}
\affiliation{Sydney Institute for Infectious Diseases, The University of Sydney, Westmead, New South Wales, 2145, Australia}

\begin{abstract}

Computational models of infectious disease can be broadly categorized into two types: individual-based (Agent-based), or compartmental models. While compartmental models can be structured to separate distinct sectors of a population, they are conceptually distinct from individual-based models in which population structure emerges from micro-scale interactions. While the conceptual distinction is straightforward, a fair comparison of the approaches is difficult to achieve. Here, we carry out such a comparison by building a set of compartmental metapopulation models from an agent-based representation of a real population. By adjusting the compartmental model to approximately match the dynamics of the Agent-based model, we identify two key qualitative properties of the individual-based dynamics which are lost upon aggregation into metapopulations. These are (1) the local depletion of susceptibility to infection, and (2) decoupling of different regional groups due to correlation between commuting behaviors and contact rates. The first of these effects is a general consequence of aggregating small, closely connected groups (i.e., families) into larger homogeneous metapopulations. The second can be interpreted as a consequence of aggregating two distinct types of individuals: school children, who travel short distances but have many potentially infectious contacts, and adults, who travel further but tend to have fewer contacts capable of transmitting infection. Our results could be generalised to other types of correlations between the characteristics of individuals and the behaviors that distinguish them.

\end{abstract}

\maketitle

\section{Introduction}


Computational modelling of infectious disease dynamics comes in two general forms: those that explicitly model each member of the population, and those that treat the population as a statistical ensemble of indistinguishable entities. The most well-established techniques are of the latter type, and compute the dynamics of susceptible and infected populations using compartmental models that specify time-dependent infection rates based on the current numbers of susceptible and infected individuals in the population. A well-known model of this type is the SIR model of Kermack and McKendrick \cite{kermack1927contribution}. Such models are advantageous due to their simple formulation and small parameter space, which facilitate the explicit evaluation of the assumptions implicit in the rate equations and make fitting the model to data straight-forward for cases in which those assumptions are approximately valid. 

There are two primary critiques of these compartmental approaches. The first relates to the assumptions made about the type of dynamics, based on average rate equations, which can be interpreted explicitly in stochastic implementations of compartmental models. These formulations assume exponentially distributed dwell times within compartments (i.e., recovery occurs at a constant average rate). In the SIR model, this assumption produces an exponentially-distributed {\it generation interval}, which defines the relationship between the basic reproductive number $R_0$ (the average number of secondary infections produced by a primary case), and the initial exponential growth rate of the infected population $r$ \cite{wallinga2007generation}. The critique is that there is a universal lack of empirical evidence for these types of dynamics in documented processes of disease transmission, with real generation interval distributions differing markedly from the exponential form. That is, the individual-level disease natural history implied by the SIR model is not justified based on any empirical evidence. More generally, observed dwell times in various disease compartments (i.e., incubation period distributions, or periods of symptom expression) are not exponentially distributed in epidemics of real-world pathogens \cite{nishiura2007early,virlogeux2015estimating,ferretti2020quantifying}. 

A second common but less fundamental critique is that compartmental models are limited in their capacity to account for heterogeneity in population properties and structure. The compartmental approach assumes each individual in a compartment has identical disease susceptibility, infectiousness, and contact frequency with others. Furthermore, in its simplest form, the approach assumes that each individual has a uniform probability of encountering any other individual in the population, an assumption which neglects social behaviours that are known to be important in the spread of infectious disease (i.e., recurrent mobility patterns and social clustering in households or schools). This critique regarding heterogeneity is less fundamental because it can be addressed with implementations that introduce added layers of complexity (additional compartments), and modulate the interaction rates between them to approximate population heterogeneity on a chosen scale. 

The most direct approach to addressing both of the above critiques is to simulate the disease natural history and population heterogeneity explicitly with Agent-based models (ABMs, also referred to frequently as individual-based models) in which each individual of the population is treated as a discrete, unique entity. While ABMs can, in principle, account for all known heterogeneity in population structure and behaviour, in practice only a small subset of known factors can be included in any given framework. The choice of which factors to include is typically constrained by the available data, required for model construction. Due to data constraints, it is difficult or impossible to faithfully incorporate all aspects of population heterogeneity that could be important for disease dynamics. On the other hand, the more factors that are included in such models, the more difficult it is to establish which of them are responsible for any observed departures from the dynamics of compartmental models. 

Over the last several decades, there has been a sustained effort in the infectious disease modelling community to understand the role of population heterogeneity. This has led to hybrid models that capture some aspects of population structure explicitly while treating others with homogeneous approximations. For example, the popular metapopulation approach treats discrete subpopulations as homogeneous entities that interact with one-another through coupling equations of various types \cite{balcan2010modeling,keeling2010individual,colizza2008epidemic,diekmann2012mathematical}. Another approach, the degree-based mean field approximation, attempts to incorporate heterogeneous contact patterns into continuum models by making assumptions about the relative time-scales of social network dynamics and infectious disease spread \cite{pastor2015epidemic}. Other work has focused on incorporating non-linear terms into compartmental frameworks to directly account for the global consequences of heterogeneity on the infectious disease dynamics \cite{bansal2007individual,connell2009comparison,stroud2006semi,novozhilov2012epidemiological}. 

While many studies have addressed these questions of heterogeneity, it is rare to find explicit comparisons between different modelling approaches (but see e.g., ~\cite{ajelli2010comparing,connell2009comparison}). In this work, we compare the results of a complex ABM with a metapopulation model (MPM) derived from the same population data. By aggregating the individuals into subpopulations, we explicitly remove demographic heterogeneity, the nuances of disease natural history progression, and the effects of social clustering. However, the MPM still accounts for medium-scale recurrent mobility patterns, and the basic transmission characteristics of the modelled pathogen (i.e., the basic reproductive ratio $R_0$, and mean generation interval). After comparing the results of the ABM with both stochastic and deterministic implementations of the MPM, we explore some generic approaches of correcting for observed discrepancies and discuss the results in the context of which factors, included in the ABM, are responsible for the qualitative differences. 

Our results indicate that incorporation of nonlinear terms into transmission equations is not sufficient to account for discrepancies related to the interplay between demographic heterogeneity and coupling strength between different subpopulations. However, when we also apply a global modulation of the coupling strengths computed for the homogeneous metapopulation model, we are able to recover important qualitative aspects of the ABM dynamics. In this case, our results indicate that behavioural heterogeneity manifests both as a weakening of transmission strength with depletion of the susceptible population, as well as a reduction in coupling strength between communities. This is likely due to the different travel behaviours of children and adults. Children, who tend to have higher susceptibility, infectiousness, and contact rates, contribute substantially less to long-range population fluxes associated with commuting behaviour. For cases in which individual properties correlate with different types of behavioural tendencies, heterogeneity will likely alter both the dynamics within local regions and the coupling between regions.  



\section{Methods}

\begin{figure}
\includegraphics[width=0.9 \textwidth]{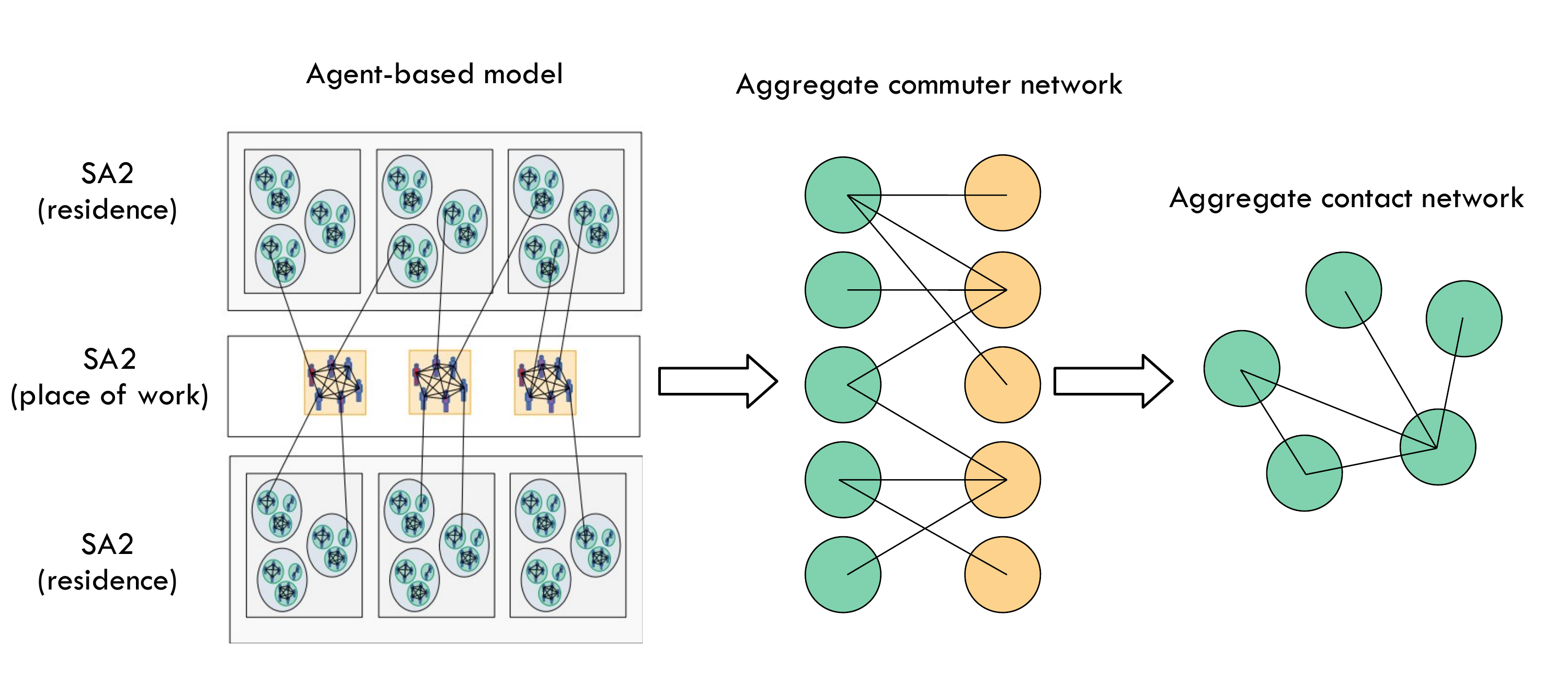}
\caption{Schematic of the aggregation procedure used to convert the detailed travel patterns in the agent-based model into the large-scale contact network used in the metapopulation model. Travel of individuals from home to workplaces or schools is aggregated into larger regions for a bipartite origin-destination matrix representation. Mass action expressions are then applied in order to generate a mixing matrix from those travel volumes.}
\label{fig_aggregation}
\end{figure}

\subsection{Agent-based model}

In general, ABMs attempt to explicitly capture the salient features of individual behaviour in order to simulate non-equilibrium dynamics on the level of heterogeneous collectives. 

The ABM of influenza transmission in Australia used in this work has been described in detail in several previous publications \cite{cliff2018investigating,zachreson2018urbanization}, so only a brief overview will be given here. The aspects of the modelling framework that are most relevant to the present study are the mid- to long-range mobility network \cite{fair2019creating} and the heterogeneity between students and adults which constitutes the most significant classification in terms of the individual behaviour captured by the model. Notably, the ABM individually simulates approximately 23M Agents, and effective use of it requires high-performance computing facilities. 

\subsubsection{Population model}

The population model is built from the Australian Census of 2016 (Australian Bureau of Statistics). It incorporates local heterogeneity and connectivity at the level of households, household clusters, neighbourhoods, and communities. Workplaces and schools provide settings for mixing outside of the local community context, and give rise to the large-scale connectivity patterns captured by the metapopulation model derived here. 

Individuals are assigned to households based on the local distribution of household sizes and compositions. The ages of individuals are assigned based on the composition of their households, which effectively reproduces the age distribution computed independently from the census \cite{cliff2018investigating}. This procedure produces a calibrated, parsimonious model of the Australian residential population. 

Agents between the ages of 18 and 65 years are assigned to workplace groups of 10 individuals based on commuting data provided in the Census \cite{fair2019creating}. Children between the ages of 5 and 18 years are assigned to schools based on proximity and capacity (they are randomly assigned to the nearest school with available capacity). School locations and capacities were determined from a dataset provided by the Australian Curriculum Assessment and Reporting Agency as described previously \cite{zachreson2018urbanization}.

\subsubsection{Disease model}

The disease model used in the ABM is designed to simulate influenza infection within an individual and consists of the following agent states: susceptible, latent (exposed, non-infectious), infected (infectious), and recovered. Additionally, 33\% of agents are asymptomatic if infected, with infectiousness reduced by a factor of 2 relative to symptomatic agents. The disease natural history model used in the ABM is described in detail in our previous work \cite{cliff2018investigating}. 

\subsubsection{Scenario}

The ABM simulates pandemic influenza in Australia, and initiates the Australian epidemic with importations (exogenous introduction of index cases) within 50 km of international airports. Relative importation rates are determined by the international arrival numbers at each airport, which produces the initial (seeding) conditions of each epidemic simulation which then proceeds through stochastic transmission of disease using binomially distributed random variables that determine the infection status of each agent, at each 12-hr step of the simulation. Each scenario terminates when the infected population reaches 0, or the pre-determined time limit is exceeded.

\subsection{Metapopulation model}

\begin{figure}
\includegraphics[width=0.9 \textwidth]{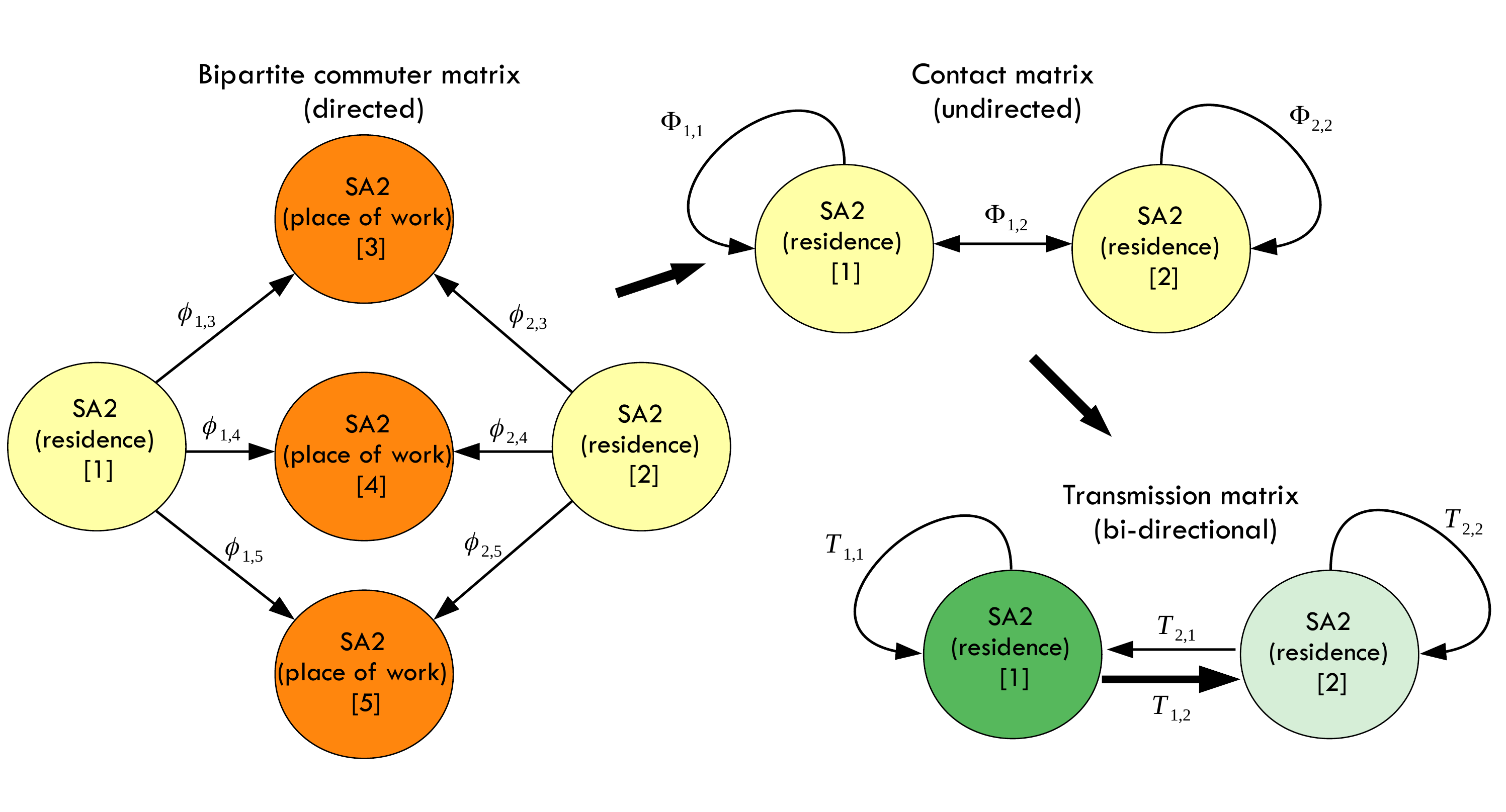}
\caption{A schematic representation of the projection process used to build the matrix of coupled metapopulations. From a bipartite commuter matrix, mass-action equations give a symmetric (undirected) contact matrix describing the strength of potential disease transmission between and within areas. Once the epidemic is seeded into this network of metapopulations, a disease transmission matrix describes the rates at which infection may spread between coupled areas with transmission computed based on the populations of infected and susceptible individuals, and the strength of the coupling between each location. }
\label{fig_projection}
\end{figure}

The goal in creating our metapopulation model is to preserve the large-scale mobility patterns of the population, while averaging over all local heterogeneity. The basic process is illustrated in Figure \ref{fig_aggregation}. We do this by systematically building up a transport matrix on the scale of statistical area level two (SA2). Each of these $n$ partitions contains approximately $1 \times 10 ^4$, to $3 \times 10 ^ 5$ individuals. In the ABM, individuals either stay within their home SA2 during the day or they commute to school or work in another area. The MPM consists of a matrix representing the mixing patterns between these locations. The first step to creating such a model is to build a bipartite adjacency matrix $\bm{\phi}$ describing commuting patterns between home and work: 
\begin{equation}
\bm{\phi}_{ij} =\frac{ \text{OD}(\text{SA2UR}_i \rightarrow \text{SA2POW}_j)}{N_i}\,,
\end{equation}
where the $ij$th entry is the fraction of individuals residing in area $i$ who commute between their home in area $i$ and their destination in area $j$, on a typical day (here, $N_i$ is the population of individuals living in area $i$). This bipartite structure is illustrated on the left hand side of Figure \ref{fig_projection}. Building this commuter matrix is a simple process of iterating through each individual in the ABM population, and incrementally increasing the entry of the commuter matrix corresponding to that individual's home-destination pair.

To convert the resulting bipartite mobility network into a coupling matrix, we apply standard methods involving simple mass-action expressions. For a given pair of locations, the coupling between them is computed as:  

\begin{equation}\label{Eq_Phiij}
\bm{\Phi}_{ij} = \sum\limits_{k = 1}^{n} \bm{\phi}_{ik} \bm{\phi}_{jk} \frac{1}{N_{k}^{p}} \,,
\end{equation}
where 
\begin{equation}
N_{j}^{p} = \sum\limits_{i = 1}^{n} \bm{\phi}_{ij}N_i\,,
\end{equation}
is the population {\it present} in location $j$ during work hours. 

The mixing matrix $\bm{\Phi}$ is a weighted, symmetric, unipartite adjacency matrix representing mixing probabilities between individuals residing in each respective location, so that multiplication of $\bm{\Phi}_{ij}$ by subpopulation numbers from areas $i$ and $j$ gives the number of contacts between those subpopulations per unit time. 

To convert this model of coupled metapopulations into a full infectious disease model, we first need to define our disease dynamics. For simplicity, generality, and approximate correspondence to the ABM, we will use the canonical SIR dynamics: 
\begin{equation}\label{eq_sir}
\frac{dS}{dt} = -\beta S I\,, ~~~
\frac{dI}{dt} = \beta S I - \gamma I \,, ~~~
\frac{dR}{dt} = \gamma I\,,
\end{equation}
in which $S$ is the proportion of susceptible individuals, $I$ is the proportion of infected individuals, and $R$ is the proportion of recovered individuals, while $\beta$ and $\gamma$ are the transmission and recovery rates, respectively. Strictly speaking, $\beta$ incorporates also the rate at which contacts capable of transmitting infection occur (in some descriptions, the contact rate is included as a separate parameter).  

Converting $\bm{\Phi}$ into a transmission matrix describing infection dynamics between subpopulations simply requires multiplication by infected and susceptible populations, a transmission rate per contact, and a contact rate per unit time (here, the these two terms are absorbed into the transmission rate $\beta$): 
\begin{equation}\label{Eq_Tij}
T_{ij} = \beta\bm{\Phi}_{ij} S_i N_i\,,
\end{equation}
which gives the rate of transmission from a single infected individual residing in area $j$ to the susceptible population of area $i$, per unit time. This framework allows us to build several different implementations of metapopulation models the results of which can then be compared to those of the ABM. Here, we discuss a deterministic discrete-time implementation, a stochastic discrete-time implementation, and a stochastic continuous-time implementation. 

\subsubsection{Deterministic, discrete time}

To compute the dynamics of the deterministic version of this metapopulation model, all that is required is the combination of the transmission and recovery terms, allowing numerical integration of the coupled rate equations. The vector of infected populations $\bm{I}$ at time $t + \Delta t$ is computed as:
\begin{equation}
\bm{I}(t + \Delta t) = [\mathbb{I} + \Delta t (T - \gamma \mathbb{I})]\bm{I}(t)\,,
\end{equation}
where $\mathbb I$ is the identity matrix of size $n$ (note that we have used $I$ to represent the fraction of infected individuals in Equation \ref{eq_sir}, while $\bm{I}$ is a vector of infected populations in each subregion). The infection numbers in each location depend on the incoming transmissions from all other areas, local transmissions, and local recovery of individuals. Since these are continuous dynamics, non-integer values of infected populations are possible. While this condition may be acceptable when the infected and susceptible populations are both large, it makes interpretation difficult in the initial stages of the epidemic, and leads to substantial discrepancies in the results of numerical simulations, as we will discuss in the results sections. 

\subsubsection{Stochastic, discrete time}

In the discrete-time stochastic description of the structured SIR dynamics, the force of infection on the subset of individuals living in region $i$ is computed from the mixing terms given in Equation \ref{Eq_Phiij}, aggregated over each location's neighbours $j$, multiplied by the number of ill individuals in each neighbouring region, the transmission rate $\beta$, and the chosen time interval $\Delta t$. This gives the individual-level probability of transitioning from the susceptible to infected state, computed independently for each SA2 subregion $i$:
\begin{equation}
p(S\rightarrow I~|~S,~\text{SA2}_i) =1 - \text{exp}({ -\beta\Delta t\, \sum_j \bm{\Phi}_{ij}\bm{I}_j}), 
\end{equation}
while recovery only depends on the rate $\gamma$ and the time interval: 
\begin{equation}
p(I\rightarrow R~|~I) = 1 - \text{exp}({-\gamma\Delta t}). 
\end{equation}

\subsubsection{Stochastic, continuous time}

To ensure that the dynamics computed by the discrete-time implementation were sufficiently free of artefacts due to time discretisation, we implemented a continuous stochastic version of the model using the Gillespie algorithm. To do so, at each time $t$ the probability of each possible next event was computed, and the next event chosen by inverse-transform sampling of the corresponding cumulative distribution function. The delay $\delta t$ between each subsequent event was computed in the standard way by sampling from the exponentially-distributed inter-event times $\delta t \sim \rho e^{-\rho}$ where $\rho$ is the sum over all individual event rates. In the system described here, there are $2n$ possible events, consisting of either recovery or infection events, in each of the $n$ subregions. By comparing the results of the continuous time model to those of the discrete-time stochastic model, we determined that a discrete time step of $\Delta t \leq 0.1~\text{d}$ was sufficient for convergence of the discrete dynamics. Because we did not identify any significant differences between the results of the continuous time simulation and those of the stochastic discrete time model, we do not report the results of the continuous time model, but have included the source code in the online Zenodo database (DOI: 10.5281/zenodo.5762581). 

\subsection{Extensions to the SIR framework}

After discussing the differences between the results produced by the ABM and those produced by a comparable MPM, we introduce two extensions to the standard SIR modelling framework. These extensions modify two aspects of the dynamics that are not captured in the MPM framework described above. The first of these is the tendency for local depletion of the susceptible population \cite{stroud2006semi}. In tightly-connected clusters of individuals such as households or workplaces, disease spread is enhanced by the higher-than-average contact rates between the individuals in the cluster. However, such clustering also produces lower-than-average transmission rates between such groups, and this leads to depletion of the local susceptible population during disease transmission. Mathematically, we attempt to capture such effects in a  phenomenological way by introducing a saturation parameter $\lambda > 0$ that decelerates transmission as the susceptible population decreases: 

\begin{equation}\label{Eq_Tij}
T_{ij} = \beta_i\bm{\Phi}_{ij} S_i\,,
\end{equation}
where
\begin{equation}\label{eq_lambda}
\beta_i = \beta [S_i]^{\lambda}\,,
\end{equation}
which produces a system in which the dynamics are initially identical to those of the base model but transmission strength decreases more quickly as the susceptible population is depleted. 

In addition to this saturation parameter, we explore the effects of altering the properties of the mixing matrix to enhance local mixing and inhibit inter-regional transmission. By doing so, transmission within regions will accelerate and transmission between regions will slow down. Our rationale for introducing such an effect is that we would like to understand the phenomena produced by correlation between decreased travel behaviour and increased social mixing behaviour of school children, without explicitly simulating the underlying heterogeneity. In the ABM, school children have larger mixing groups (classroom groups, grade groups, and school groups), but typically remain local to their home region during the day. To modify our metapopulation model in order to adjust for these effects, we alter the commuter flows described in the OD matrix $\phi$, by uniformly moving commuters from off-diagonal components into the diagonal component (those who remain in their local areas). To manipulate the degree to which this occurs, we introduce the parameter $\sigma$, which modulates the original OD matrix as follows: 
\begin{align}\label{eq_sigma}
&\tilde{\phi}_{ij} = \phi_{ij}(1 - \sigma)\,, ~i \neq j \,,\\ \nonumber
&\tilde{\phi}_{ii} = \phi_{ii} + \sum\limits_{i\neq j} \sigma \phi_{ij}\,,
\end{align}
so that a fraction $\sigma$ of travelers remain within their local regions instead of transiting outside of them. These traveler flows are used to compute the mixing matrix $\bf{\Phi}$, which determines the transmission strength between regions. 

With these two degrees of freedom added to the base MPM configuration, we can tune the relatively simple MPM model to closely match the dynamics produced by the complex ABM.

\subsection{Matching the models}

There are several fundamental differences between the ABM and the MPM. These relate to the details of disease natural history (i.e., time-dependent recovery probabilities in the ABM), heterogeneity of individual behaviours which determine structured mixing patterns within SA2 regions (which are considered as homogeneous populations in the metapopulation framework), and social clustering across spatial regions due to habitual contact patterns between travellers (i.e., the interactions of co-workers). 

In order to approach the question of how these fundamental factors affect the results of simulations, we first calibrate the two frameworks to match important quantities that must be conserved across models to facilitate a fair comparison.     

\subsubsection{The basic reproductive number $R_0$, generation time, and initial growth rate $r$}

\begin{figure}[t]
\includegraphics[width=0.8\textwidth]{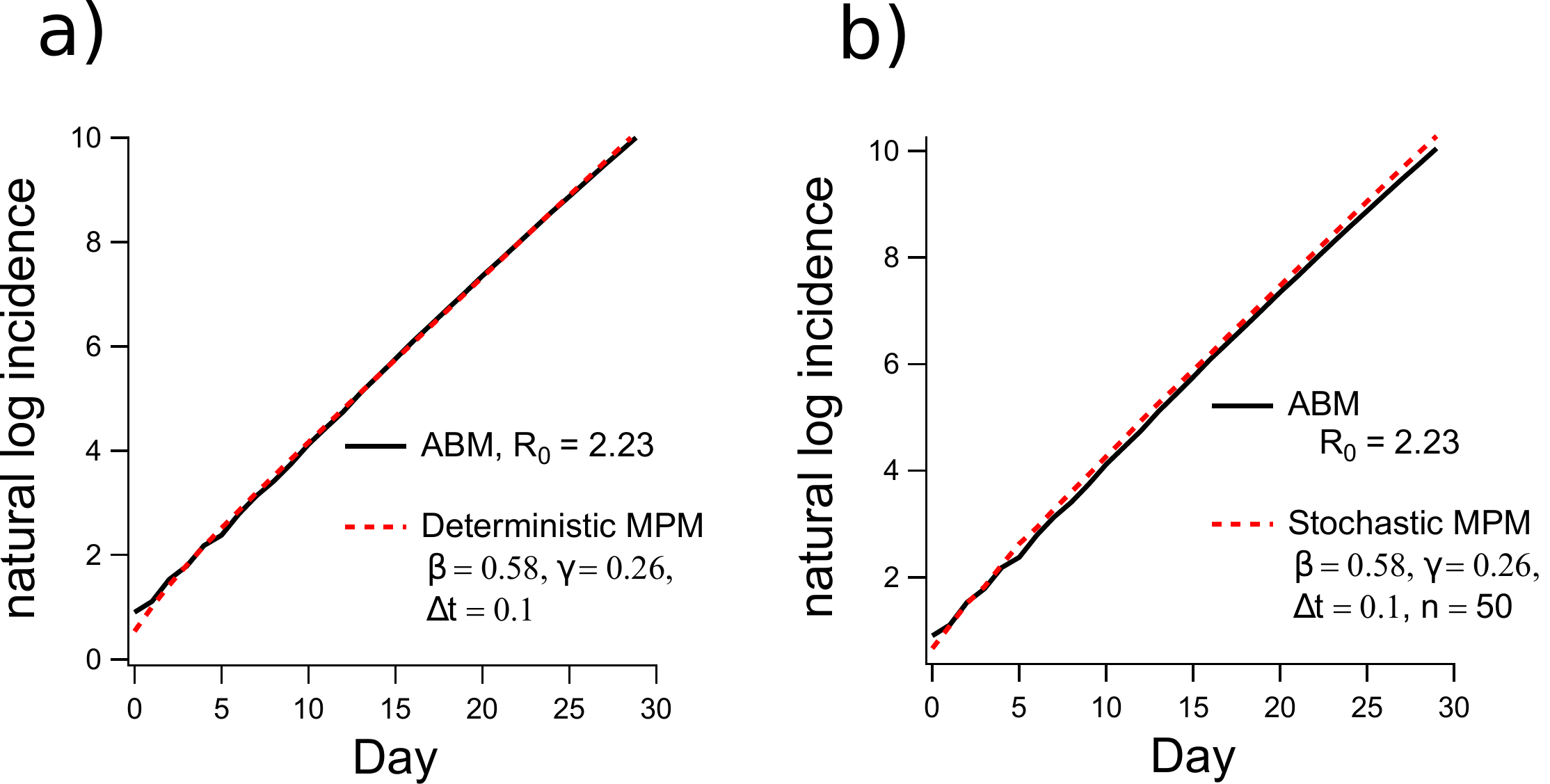}
\caption{Calibration of the initial conditions of the metapopulation models to match those of the Agent-based models. The plot shows the natural logarithm of incidence (number of new illnesses) counted on each day of the simulation. By calibrating the recovery rate of the SIR model to the generation time expected to match the exponential growth rate observed in the ABM, the initial growth of the incidence rate in the ABM is well-matched to both the deterministic and stochastic implementations of the metapopulation model.}
\label{fig_initial_growth}
\end{figure}

An essential quantity describing any model of contagion spread is the basic reproductive number $R_0$, the average number of secondary cases produced by a typical primary infection (index case), in a completely susceptible population. For the ABM, this quantity is difficult to compute analytically due to the complex nature of the population structure and disease natural history model. However, it is straight-forward to estimate $R_0$ from simulations. This allows calibration with respect to the single independent parameter $\kappa$, the transmission multiplier, in order to approximate a set value of $R_0$. We used this technique to calibrate the ABM so that $R_0 = 2.23$, that is, for typical index cases in an entirely susceptible population, there occur 2.23 secondary infections on average. The calibration procedure is described in detail in our previous paper, and takes into account the age-structured distribution of primary infections \cite{zachreson2020interfering}. In the ABM, the duration of illness is 5.5 days, with a constant 1-day latency period applied to all infections. The resulting generation time, $T_g$, the average time for the infection to propagate between infectious and susceptible individuals, was computed as $T_g\approx 3.4$ days (the ensemble of index cases and transmission events used to generate our estimate of $R_0$ was also used to estimate generation time). In SIR dynamics, $T_g$ is related to the reproductive ratio and the initial exponential growth rate $r$ as: 
\begin{equation}\label{Eq_Tc}
T_g = \frac{R_0 - 1}{r}\,.
\end{equation}
An average growth rate $r = 0.32~\text{d}^{-1}$ was estimated by calculating the mean of the incidence from the first 30 simulation days of 30 independent realisations of the ABM, and fitting to an exponential growth function. Using this growth rate, and the estimated value $R_0 = 2.23$, Equation \ref{Eq_Tc} gives $T_g = 3.84$ days, which is close to the generation time estimated from the ABM results. In the SIR dynamics, the inverse generation time $1/T_g$ is equivalent to the recovery rate $\gamma$. In the SIR model, $R_0 = \beta / \gamma$, so initialising the SIR metapopulation model with $\gamma = 0.26~d^{-1}$ and $\beta = 0.58~\text{d}^{-1}$ should give an initial dynamics approximately equivalent to those of the ABM. These parameters indeed produce almost identical initial growth dynamics to the average over ABM trajectories, in both deterministic and stochastic MPM implementations, as shown in Figure \ref{fig_initial_growth}.

\subsubsection{Initial conditions}
The results of infectious disease models in heterogeneous populations can be sensitive to the distribution of index cases. We took the following steps to ensure a fair comparison between models. In the ABM, the international pandemic scenario corresponds to a steady influx of index cases near international airports. Specifically, each international arrival produces a $0.004\%$ chance of a new index case occurring within each member of the susceptible population residing in the SA2 regions within 50~km of the corresponding airport. This incoming force of infection is continuously applied throughout the simulation. Mimicking these conditions in the metapopulation framework is straight-forward, because the seeding procedure is carried out on the same scale (SA2) in the ABM. In the discrete-time stochastic implementation, the procedure is identical to that of the ABM, while in the deterministic implementation the incoming force of infection is interpreted as a local rate of infection, applied in addition to the transmission terms described by Equation \ref{Eq_Tij}. In the continuous-time stochastic implementation, the seeding rates are added to the transmission and recovery rates when computing the global event rate, and appended to the infection rate terms for each of the affected SA2 regions when computing the next-event CDF. 
After matching the initial conditions and growth dynamics, the results of the ABM can be compared to those of the MPM.  [Note: the 1-day latency period and stochastic symptom expression in the ABM produces an offset of 3 days between the incidence counts computed by the ABM and those computed for the MPM, which does not account for progressive symptom expression. In the results below, the ABM data has been shifted by this interval to facilitate visual comparison.]

\FloatBarrier

\section{Results}

\begin{figure}
\includegraphics[width= \textwidth]{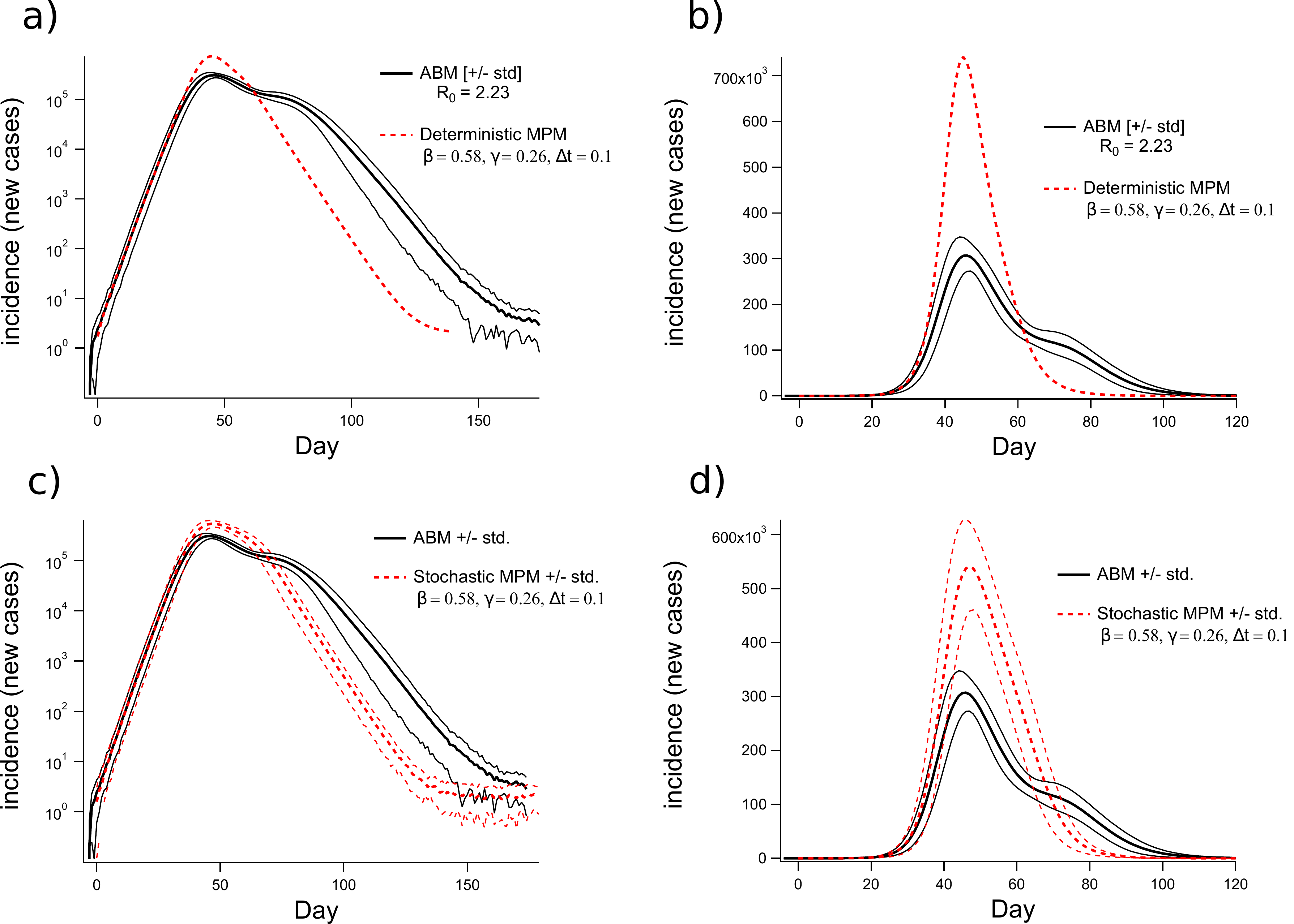}
\caption{Epidemic incidence curves comparing the results of the Agent-based model with those of the deterministic and stochastic implementations of the metapopulation model. Black curves correspond to the ABM results (upper and lower curves show $\pm$ std.). Red curves in subplots (a) and (b) correspond to log-scaled and linearly scaled incidence (respectively) for the deterministic implementation, while red curves in subplots (c) and (d) show log-scaled and linearly scaled incidence computed by the stochastic implementation. For the stochastic implementation, the average and standard devaition were computed from 50 independent runs.}
\label{fig:ABM_vs_MBM_base}
\end{figure}

Without correcting for population heterogeneity, the metapopulation model calibrated to the same initial growth rate and initial conditions as the ABM overestimates peak prevalence and underestimates epidemic duration. This general finding is true for both the deterministic and stochastic implementations of the MPM. Of the two implementations, the stochastic MPM is qualitatively more consistent with the ABM (Figure \ref{fig:ABM_vs_MBM_base}).

\begin{figure}[t]
\includegraphics[width= \textwidth]{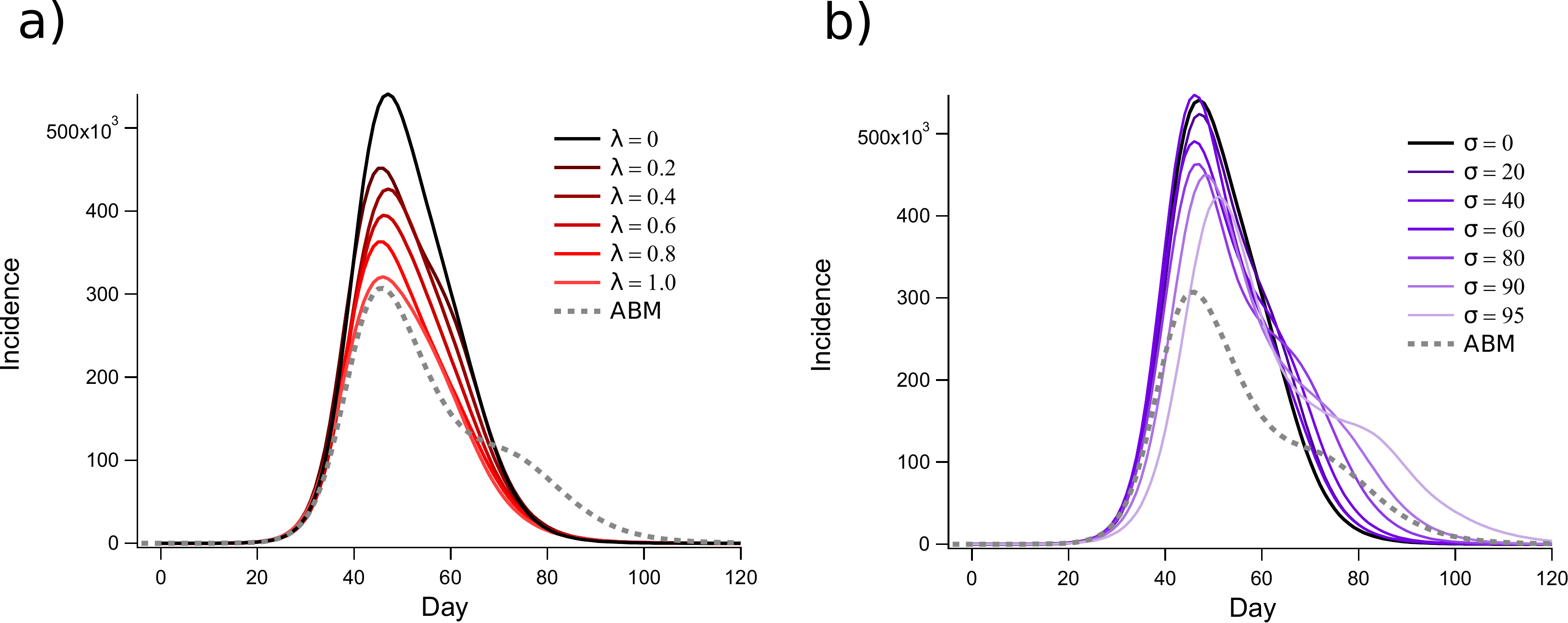}
\caption{Incidence curves produced by independently varying (a) the susceptibility attenuation parameter $\lambda$, and (b) the inter-community interaction parameter $\sigma$, in the extended stochastic metapopulation model.}
\label{fig:lambda_sigma}
\end{figure}

To account for the effects of heterogeneity, Equation \ref{eq_lambda} and Equation \ref{eq_sigma} introduce the correction factors $\lambda$ and $\sigma$, respectively. Varying the transmission attenuation parameter $\lambda$ decreases peak incidence without altering the initial growth rate, as shown in Figure \ref{fig:lambda_sigma}(a). On the other hand, altering the inter-region mixing parameter $\sigma$ marginally reduces peak incidence and slightly increases peak timing, while substantially broadening the incidence curve and, for large enough values ($\sigma \approx 0.9$), produces observable bi-modality, a feature of the ABM produced by weak connections between urban and regional areas, and discussed at length in our previous work (Figure \ref{fig:lambda_sigma}b) \cite{zachreson2018urbanization}. 

\begin{figure}[t]
\includegraphics[width=0.95\textwidth]{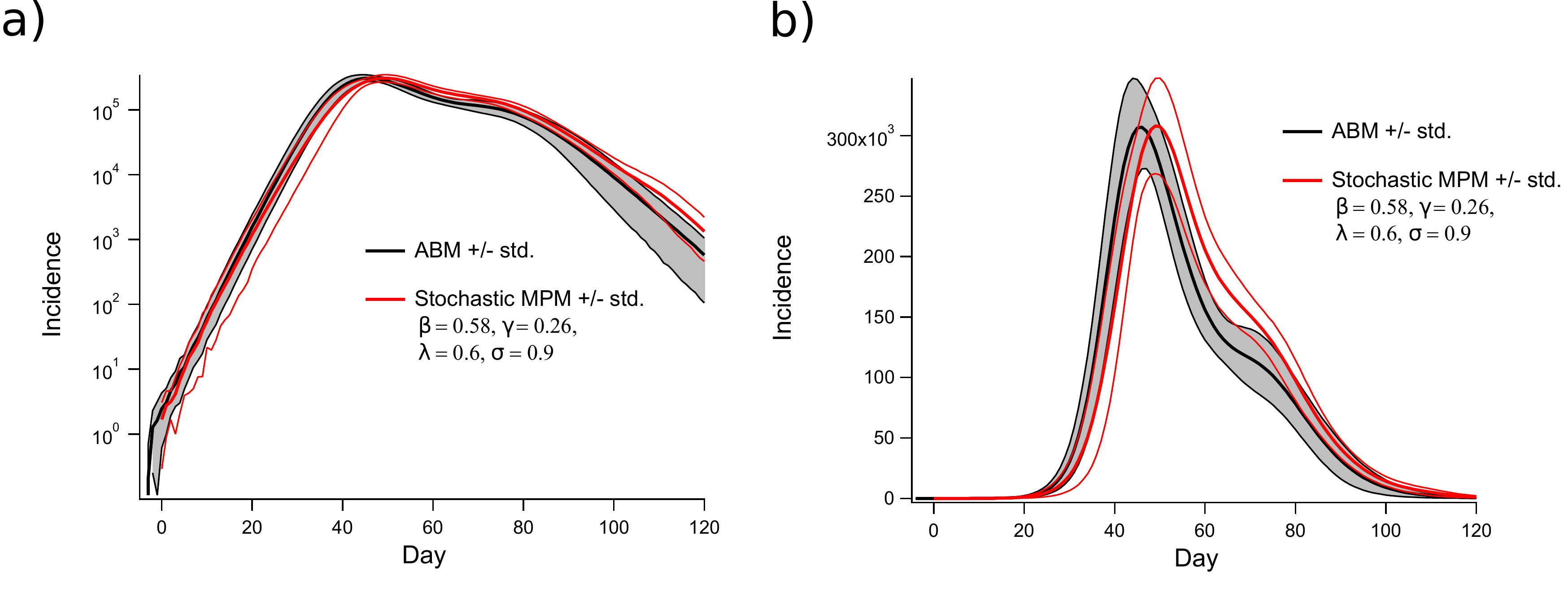}
\caption{Log-scaled (a) and linear-scaled (b) incidence curves comparing the ABM with the extended stochastic MPM after fine-tuning the saturation parameter $\lambda$ and inter-community interaction parameter $\sigma$ to qualitatively match the ABM results. Initial growth rates and peak incidence are very closely matched between the models. Peak timing is slightly over-estimated by the adjusted MPM.}
\label{fig:lambda_sigma_ABM_match}
\end{figure}

By tuning the attenuation and inter-region mixing parameters, the mean incidence trajectories of the MPM can be made to closely match those of the ABM (Figure \ref{fig:lambda_sigma_ABM_match}). For attenuation $\lambda = 0.6$ and inter-region mixing reduction of $\sigma = 0.9$, the growth rate, peak incidence (and its run-to-run standard deviation), and the bimodal character of the ABM are reproduced in the MPM implementation. The inter-region mixing parameter $\sigma$ produces a delay in the peak incidence. This is likely due to the uniform alteration in diagonal vs. off-diagonal travel matrix elements, over all regions. 

\begin{figure}
\includegraphics[width=\textwidth]{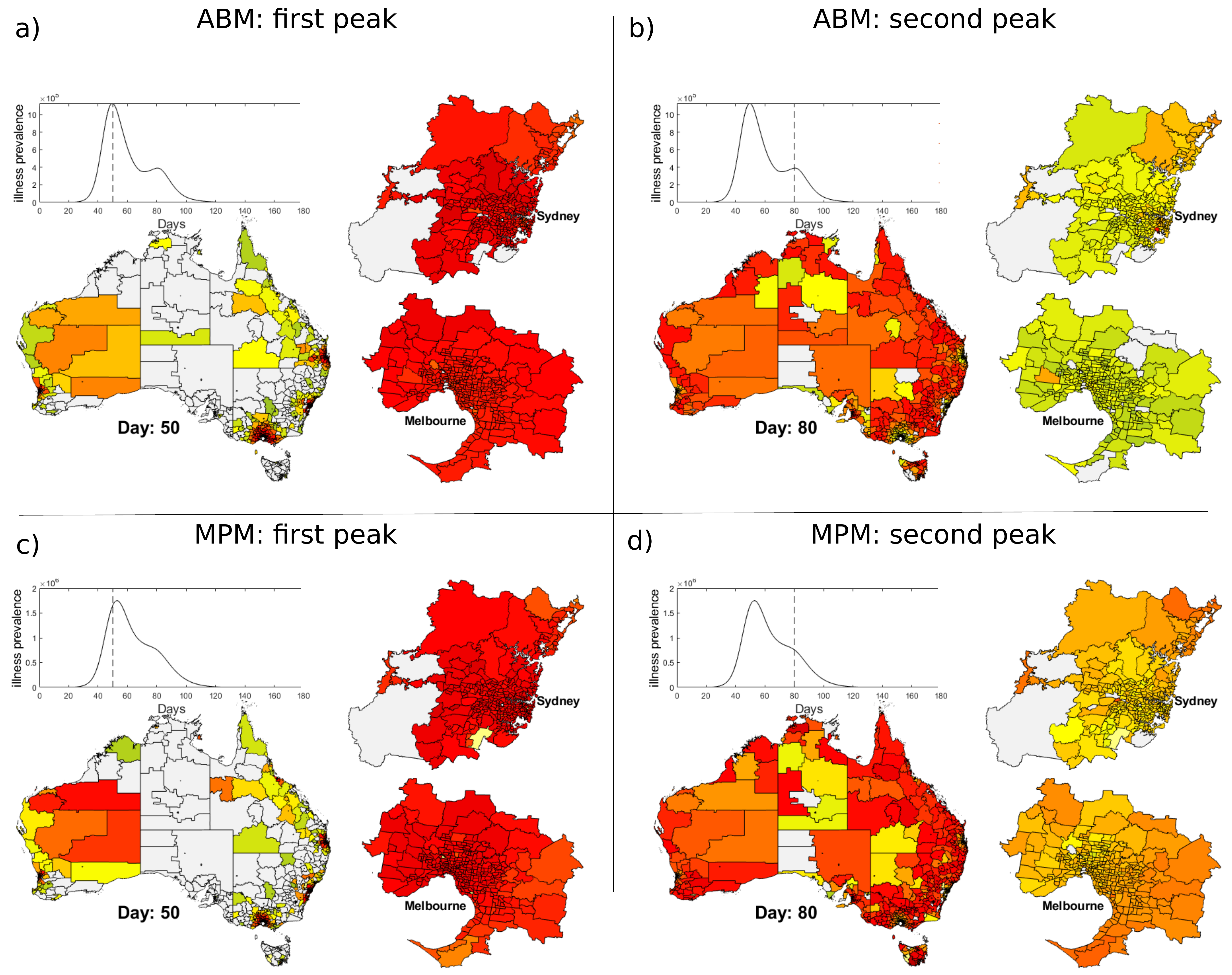}
\caption{Qualitative comparison of spatiotemporal transmission patterns generated by the ABM (a,b) and MPM (c,d). Choropleth maps of disease prevalence show the spatial distribution of illness prevalence from single simulation instances, these were selected at random from the ensemble of simulations conducted for this work, to demonstrate qualitative aspects of the dynamics produced. The selected snapshots from days 50 and 80 of each simulation show the distribution of illness during the first prevalence peak (a, c), and the second peak (b, d), respectively. The MPM results in (c) and (d) were generated using the tuned parameters $\sigma = 0.9,~ \lambda = 0.6$.}
\label{fig:maps}
\end{figure}

Finally, to investigate the qualitative correspondence between the dynamics produced by the ABM and adjusted MPM, we visualised the spatiotemporal distribution of illness prevalence as computed by both models. Examples are shown in Figure \ref{fig:maps}. In both models, outbreaks begin in densely-populated urban areas, where international arrivals are concentrated and epidemic seeding is more likely. This leads to an initial peak concentrated in the larger urban centres, followed by a later second peak corresponding to urban regions and smaller urban areas with lower levels of international air traffic. As expected based on the correspondence shown in Figure \ref{fig:lambda_sigma_ABM_match}, the dynamics of the ABM are born out in the adjusted MPM.

\section{Discussion}

A comparison between an agent-based and a structured metapopulation stochastic model, using a pandemic event in Italy simulated within the GLobal Epidemic and Mobility (GLEaM) framework~\cite{ajelli2010comparing}, pointed out that the MPM consistently yielded a larger incidence than the ABM.  This bias was explained by the assumption of homogeneity utilised in the MPM, as well as different structures in the intra-population contact patterns captured by the respective approaches. For example, the largest discrepancy between the two models was reported for the 60+ age class --- a ``class with the most marked difference in household structure and workplace habits that cannot be taken into consideration in the metapopulation level''~\cite{ajelli2010comparing}. Our comparative analysis also indicated that aggregation of heterogeneity in household composition and social mixing behaviour produced a similar result. These discrepancies were reported to increase with higher values of the basic reproductive number $R_0$~\cite{ajelli2010comparing}, and hence, this tendency forms an elevated concern in modelling more infectious diseases, such as the COVID-19. Indeed, pandemic spread of SARS-CoV-2, and its more infectious Delta (B.1.617.2) variant, was often driven by structural factors, especially in areas characterised by socioeconomic disadvantage profiles, with high-density housing, multi-generational and shared households, higher concentrations of essential workers~\cite{chang2021nowcasting}, increasing the risks of exposure within overcrowded living spaces~\cite{rasul2021socio}, and disproportionately affecting rural and urban areas~\cite{aneja2021assessment}.
 
Agent-based models manifest both behavioural and mechanical adequacy of the disease transmission mechanism~\cite{glennan2005modeling,maziarz2020agent}. While ``behavioural adequacy'' ensures that simulated epidemic patterns agree with the observations, ``mechanical adequacy'' guarantees a concordant natural history of the disease within an individual, representing stages of the pathogenesis from exposure and incubation to the infectivity peak and then to recovery or death. In a large-scale ABM, each infected agent follows a profile which is stochastically generated from an explicitly defined distribution, thus intrinsically representing diversity of possible histories. When the natural history of disease exhibits non-trivial dependencies (e.g., with respect to pre-symptomatic or asymptomatic infectivity), an aggregation of individual agent profiles within a MPM is likely to average out nuances of the time-dependent infectiousness profile. Aggregation of these individual-level factors impedes the design of intricate targeted interventions based on nuanced timing considerations (e.g., the isolation of infectious contacts before they can infect others). 
 
In general, one may need to consider a balance between (i) increased realism of agent-based models, achieved at the cost of a more laborious calibration of internal parameters~\cite{maziarz2020agent,hoertel2020stochastic} and a detailed reconstruction of mobility patterns~\cite{fair2019creating}, and (ii) coarse-grained compartmental approximations of the overall epidemic dynamics, attained with substantially reduced design, calibration, and computational costs. To re-iterate, there is much promise for hybrid models which enrich homogeneous approximations with some aspects of population heterogeneity~\cite{arino2003multi,belik2011natural}, or that utilise individual-based models to define sensitive initial conditions \cite{zachreson2021covid}. 
 
However, modelling targeted interventions remains a challenge even for these hybrid approaches, since the intensity of interactions within mixing groups dynamically changes in response to various spatiotemporal constraints (social distancing, vaccinations, school closures, etc.). These fine-grained changes may generate local and transient effects that can be lost during an aggregation, complicating modelling of intervention efforts focused on super-spreading events~\cite{endo2020estimating}, epidemic risk assessment for vulnerable communities~\cite{pluchino2021novel}, and estimation of age-dependent hospitalisation and fatality rates~\cite{nyberg2021risk,levin2020assessing}.

This work demonstrated that by introducing two global correction factors, for a total of four free parameters ($\beta, ~\gamma, ~\sigma, ~\lambda$), a simple metapopulation model can be tuned to closely approximate the dynamics observed from a complex ABM tailored to simulate pandemic influenza H1N1 in the Australian population. The base MPM was derived from a detailed accounting of mobility between regions, to match the mixing patterns captured by the ABM. To match the ABM results, the base MPM was corrected to globally attenuate inter-region transmission, and account for local depletion of susceptible populations. The final, adjusted MPM combines a detailed accounting of mobility with a phenomenological model of disease transmission. This approach balances a key advantage of the ABM (detailed mobility patterns), with the capacity for efficient calibration and low computational cost. 

In the ABM, the demographic properties of each area influence the transmission strength between pairs of regions, which suggests that a more accurate correction procedure should take into account the proportions of different sub-populations in each region (i.e., children and adults). If these demographic distributions were taken into account explicitly, the phenomenological correction factor $\sigma$ could be replaced element-wise with a data-derived $n \times n$ correction matrix, modifying transmission strength between each pair of regions based on their local demographics. On the other hand, the MPM developed here requires only mobility information and global observations of disease incidence for its formulation, and could be a valuable approach when demographic information (and its relationship to disease spread) is uncertain or unavailable. The MPM formulation here uses ``perfect" mobility data, for a one-to-one match with the ground truth as specified by the ABM. In real outbreaks, mobility patterns are uncertain, but can be inferred from various sources on scales of aggregation appropriate for maintaining the privacy of individuals. Future applications of the model developed here could import near real-time mobility and case data, to account for sensitive initial conditions and population mixing patterns that change dynamically during an epidemic \cite{zachreson2021risk,gauvin2020socioeconomic}. 

A key limitation to application of the MPM approach demonstrated here to forecasting of epidemics is the late divergence of the models. The calibrated base MPM closely reproduced the early dynamics of the ABM, but dramatically over-estimated peak epidemic prevalence (which translates to peak public health burden). Therefore, it remains unclear how to prospectively calibrate the correction factors $\sigma$ and $\lambda$ early in an epidemic, during the exponential growth phase when intervention decisions need to be made. To do so, a clearer understanding of the behavioural phenomena underlying these corrections is required. Unlike previous explorations that have introduced similar corrections \cite{novozhilov2008spread,novozhilov2012epidemiological,bansal2007individual,stroud2006semi}, this study used a detailed ABM as ground-truth. This simulated ground-truth offers the advantage that aspects of heterogeneity accounted for by the correction factors are known. The disadvantage of using simulated ground-truth is that other potentially important factors, such as risk-driven behavioural feedback, were not incorporated into the ABM. Therefore, it is unclear whether the correction factors introduced here would be sufficient to capture the effects of such phenomena. What our work demonstrates clearly is that salient heterogeneity can be accounted for through simple, phenomenological corrections to traditional modelling approaches. As part of a hybrid, multi-scale modelling framework, these types of methods may prove useful in forecasting epidemic dynamics and efficiently providing evidence for or against intervention policies during future epidemics.


\section{Data availability}
All simulated data, simulation code, and processing scripts necessary to reproduce the results reported here (including the MPM source code and inputs produced by the ABM population generator) are freely available and may be found on the associated Zenodo database (DOI: 10.5281/zenodo.5762581). The source code and input data used for the agent-based model population generator and epidemic simulations (ACEMod) is available in a separate database (DOI: 10.5281/zenodo.5773908).

\section{Acknowledgements}
This work was partially supported by the Australian Research Council grant DP200103005 (MP and SC). CZ is supported in part by National Health and Medical Research Council project grant APP1165876. The agent-based model used in this work (ACEMod)
is registered under The University of Sydney’s invention disclosure CDIP ref. 2019–123. 

\bibliography{references.bib}

\end{document}